\documentclass[fontsize=11pt,twoside=semi,numbers=endperiod]{scrartcl}
\KOMAoptions{headinclude=true,footinclude=false,index=toc}
\usepackage[paperwidth=150mm,paperheight=179.35mm,width=110mm,textheight=153mm,headheight=9.5mm,headsep=3mm,top=13mm,inner=20mm]{geometry} 

\usepackage{amsmath}
\usepackage{amssymb}

\usepackage{lmodern} 
\usepackage{pifont} 
\usepackage{nicefrac} 

\usepackage[ansinew]{inputenc}
\usepackage[T1]{fontenc}
\usepackage[english,ngerman]{babel}
\usepackage{textcomp} 
\usepackage[pdftex]{color,graphicx} %
\usepackage[rel]{overpic} 
\usepackage{bbm}
\usepackage{cite} 
\usepackage[german,english]{varioref} 
\usepackage{ellipsis} 
\usepackage{microtype} 
\usepackage{scrpage2}
\usepackage[%
	pdftex,
	colorlinks,citecolor=blue,filecolor=blue,linkcolor=blue,urlcolor=blue,%
	bookmarks=true,%
	bookmarksopen=false,%
	bookmarksnumbered=true,%
	pdfpagemode=UseNone,
    pdfstartview={XYZ null null 1.00},
	pdffitwindow=true,
    pdfview={XYZ null null null},
    pdfpagelayout=SinglePage,
    pdfdisplaydoctitle=true, 
    plainpages=false,
    pdfpagelabels=true,
	pdfsubject={Bell's inequality},
	pdfkeywords={Bell's inequality,separability,non-locality,hidden variables,realism},
	pdftitle={Bell's Inequality},
    pdfauthor={Astrophysical Institute Neunhof}]
	{hyperref}

\setkomafont{disposition}{\rmfamily} 
\addtokomafont{section}{\large\bfseries} 
\addtokomafont{subsection}{\normalsize\bfseries} 
\addtokomafont{pagenumber}{\LARGE}
\addtokomafont{caption}{\small}

\deffootnote[1em]{1em}{1em}{\textsuperscript{\thefootnotemark}\ }

\newenvironment{ggitemize}{\vspace{0.2\baselineskip} \begin{changemargin}{0em}{2em}}{\vspace{-0.8\baselineskip} \end{changemargin}}

\newcommand{\ggitem}[2][$\ast $]{\makebox[0em][r]{#1\ \ }{#2}\newline }%

\newcommand{\ggie}{\mbox{i.\,e.}\ } 
\newcommand{\ggeg}{\mbox{e.\,g.}\ } 
 
\newcommand{\bz}[1]{\nolinebreak\hspace{0em}\nolinebreak{}#1\hspace{0em}}
\newcommand{\al}{\iflanguage{english}{``\nolinebreak\hspace{0em}}{\glqq\nolinebreak\hspace{0.1em}}\nolinebreak}
\newcommand{\ar}{\nolinebreak\iflanguage{english}{\hspace{0em}\nolinebreak{}''\ }{\hspace{-0.45em}\nolinebreak\grqq\thickspace}}
\newcommand{\arp}{\nolinebreak\iflanguage{english}{\hspace{0em}\nolinebreak{}''}{\grqq}}

\newcommand{\mklammerp}{\mbox{\hspace{.4em}\raisebox{1.1mm}[0mm][0mm]{$-$}\hspace{-1em}\raisebox{-.4mm}[0mm][0mm]{\scalebox{.8}[.55]{$($}\scalebox{.8}{$+$}\scalebox{.8}[.55]{$)$}}\hspace{.2em}}}
\newcommand{\pklammerm}{\mbox{\hspace{.4em}\raisebox{1.1mm}[0mm][0mm]{\scalebox{.8}{$+$}}\hspace{-1em}\raisebox{-.4mm}[0mm][0mm]{\scalebox{.8}[.55]{$($}\raisebox{-.4mm}[0mm][0mm]{$-$}\scalebox{.8}[.55]{)}}\hspace{.2em}}}

\DeclareMathOperator{\dif }{d}

\definecolor{blassorange}{rgb}{1,.95,.8}
\definecolor{lila}{rgb}{.7,0,.8}
\definecolor{dkllila}{rgb}{.3,0,.7}
\definecolor{dklrot}{rgb}{0.95,0,0}
\definecolor{dklgrn}{rgb}{0,0.8,0}
\definecolor{ddklgrn}{rgb}{0,0.4,0}
\definecolor{dklblau}{rgb}{0,0,0.8}
\definecolor{Diagramm10}{rgb}{1,.58,.055}

\clearscrheadfoot 
\rohead[\vspace{-2mm}\par\pagemark ]{\vspace{-2mm}\par\pagemark}
\lehead[\vspace{-2mm}\par\pagemark ]{\vspace{-2mm}\par\pagemark}
\rehead{ } 
\lohead{ } 

\begin{document}
\selectlanguage{english}
\thispagestyle{empty} 
\raggedbottom  
\vspace*{2mm}
\centerline{\bfseries\Large What exactly is proved by the}
\vspace*{2mm}
\centerline{\bfseries\Large violation of Bell's inequality?}
\vspace*{3mm}
\centerline{{\bfseries Gerold Gr\"{u}ndler}\,\footnote{\href{mailto:gerold.gruendler@astrophys-neunhof.de}{email:\ gerold.gruendler@astrophys-neunhof.de}}}
\vspace*{1mm}
\centerline{\small Astrophysical Institute Neunhof, N\"{u}rnberg, Germany} 
\vspace*{7mm}
\noindent\parbox{\textwidth}{\small\hspace{1em}Bell's inequality has been derived several times from quite different basic assumptions, which imply different conclusions. This resulted into widespread confusion regarding the exact implications of the experimental violations of the inequality. In this article, the structures of Bell's and of Peres' derivations are analyzed, and the title question is explicitly answered.} 
\vspace*{3mm}

\section{Introduction}
Although Bell published his inequality more than 50 years ago, and although it's violation has been observed in experiments since more than 30 years, there still exists an amazing amount of confusion with regard to the \emph{exact} meaning of this finding. Does it exclude hidden variables? Does it imply non\bz{-}locality? Does it imply both, or neither of them? The confusion is mainly caused by the circumstance, that Bell's inequality is part of a `proof by contradiction', \ggie  the inequality is derived from some basic assumptions, and it's violation proves that at least one of the basic assumptions must be wrong. Actually Bell (respectively Clauser, Horn, Shimony, and Holt), Wigner, and Peres derived the inequalities, which are experimentally violated, from quite different basic assumptions. Consequently the violations of these inequalities imply quite different conclusions with respect to the different derivations. 

In this article I will 
\vspace{-2ex}\begin{ggitemize}
\ggitem{clarify the structure and the basic assumptions of the derivations by Bell (respectively Clauser, Horn, Shimony, and Holt) and by Peres,}
\ggitem{demonstrate that the derivation due to Peres implies significantly stronger conclusions than Bell's own derivation, and}
\ggitem{answer the title question of this article by an explicit list of five conclusions \hyperlink{C2Peres}{C2}$_\text{Peres}$\,\dots\,\hyperlink{C6Peres}{C6}$_\text{Peres}$\,, which can be drawn from the experimentally confirmed violation of Bell's inequality.}
\end{ggitemize}

\section{Entangled bipartite quantum systems}
Einstein, Podolski, and Rosen\!\cite{Einstein:EPR} used the correlations of entangled bipartite quantum systems, to (allegedly) prove that quantum theory is incomplete. 
\begin{figure}[!b]
\parbox{\linewidth}{\begin{overpic}{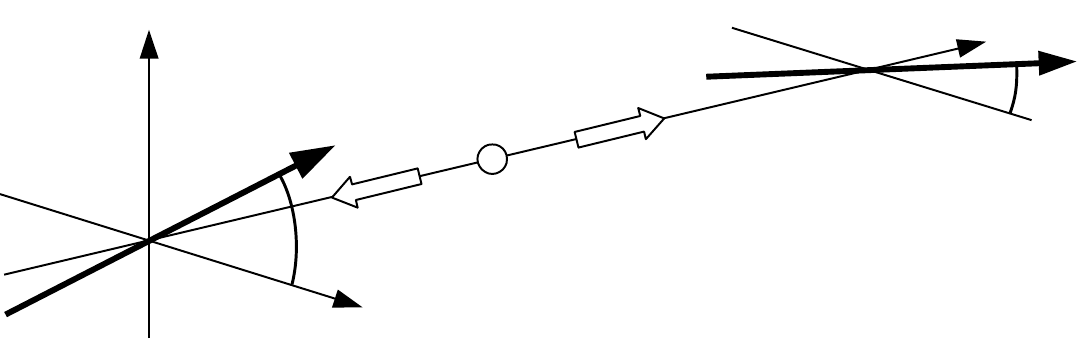}
\put(55,21.5){\small\textbf{G}}
\put(43,19.5){\small\textbf{DG}}
\put(34.7,16.7){\small\textbf{D}}
\put(28,8.8){$\delta$}
\put(94.6,22.5){$\gamma$}
\put(14.6,28){\small $x$}
\put(31.3,1){\small $y$}
\put(89,29){\small $z$}
\put(95,27.6){\small\textbf{M}}
\put(26,19){\small\textbf{M}}
\end{overpic}}%
\refstepcounter{figure}\hspace*{-.6\linewidth}\parbox[t]{.6\linewidth}{\small\vspace*{3mm}\par\hspace*{-.2mm}Fig.\,\thefigure :\ \ The singlet particle DG decays into two spin\bz{-}1/2 particles D and G\,. The spin projections of D and G are measured by Stern\bz{-}Gerlach magnets, which are rotated by $\delta $ resp.\ $\gamma $ versus the $y$ axis.}\label{fig:Setup}\end{figure} 
While EPR considered a bipartite system with canonically conjugate continuous parameters $p$ and $q$ respectively $P$ and $Q$, Bohm\!\cite{Bohm:quantmech} made the argument better amenable to experimental analysis by considering bipartite systems with each of the two parts having a discrete two\bz{-}valued spectrum. As an explicit example, he analyzed a particle DG in a singlet state, which decays into \mbox{spin\,\nicefrac{1}{2}} fragments D and G\,. The fragments are analyzed by Stern\bz{-}Gerlach magnets set to angles $\delta $ and $\gamma $ as illustrated in fig.\;\ref{fig:Setup}\,. 

As long as the particles D and G have not yet reached the magnets, quantum theory describes the spin state of the D\,\&\,G system (besides possible phase factors, which are of no relevance for the present evaluation) by the entangled state vector 
\begin{subequations}\label{mjsdngdnsd}\begin{align}
|\text{D\&G}\rangle =\sqrt{\frac{1}{2}}\,\Big(\, |\downarrow\,\rangle _\text{D}\, |\uparrow\,\rangle _\text{G}+|\uparrow\,\rangle _\text{D}\, |\downarrow\,\rangle _\text{G}\,\Big)\ .\label{mjsmngnsdc}
\end{align} 
Quantum theory doesn't assign any state vector to fragment D, nor any state vector to fragment G. It only assigns the entangled state vector \eqref{mjsmngnsdc} to the overall system. 

For arbitrary settings $\delta $ of one magnet, \eqref{mjsmngnsdc} implies a strictly anticorrelated result if $\gamma =\delta $\,, and a strictly correlated result if $\gamma =\delta +\pi $\,. EPR concluded: Either Nature must arrange non\bz{-}locally for the appropriate correlations in space\bz{-}like separated measurements, or the spin projections of the two fragments onto arbitrary space axes were determined already since the DE particle decayed. Only the second explanation seemed acceptable to EPR. As quantum theory misses to assign (before the measurements) definite spin projections onto arbitrary space axes to particles D and G, but merely supplies the entangled vector \eqref{mjsmngnsdc} of the overall system, EPR deemed quantum theory incomplete. 

The two entangled \mbox{spin\,\nicefrac{1}{2}} particles might as well be evaluated in the correlated state 
\begin{align}
|\text{D\&G}\rangle =\sqrt{\frac{1}{2}}\,\Big(\, |\uparrow\,\rangle _\text{D}\, |\uparrow\,\rangle _\text{G}+|\downarrow\,\rangle _\text{D}\, |\downarrow\,\rangle _\text{G}\,\Big)\ ,\label{mjsmngnsdd}
\end{align}
in contrast to the anticorrelated state \eqref{mjsmngnsdc}. And in many experiments, photon polarizations are evaluated instead of spin projections, with two photons D and G prepared in the correlated entangled state 
\begin{align}
|\text{D\&G}\rangle =\sqrt{\frac{1}{2}}\,\Big(\, |\text{H}\rangle _\text{D}\, |\text{H}\rangle _\text{G}+|\text{V}\rangle _\text{D}\, |\text{V}\rangle _\text{G}\,\Big)\label{msnmgnsdga}
\end{align} 
or in the anticorrelated entangled state 
\begin{align}
|\text{D\&G}\rangle =\sqrt{\frac{1}{2}}\,\Big(\, |\text{H}\rangle _\text{D}\, |\text{V}\rangle _\text{G}+|\text{V}\rangle _\text{D}\, |\text{H}\rangle _\text{G}\,\Big)\ ,\label{msnmgnsdgb}
\end{align}\end{subequations} 
with H encoding horizontal polarization, and V encoding vertical polarization. 
%

Due to the definitions 
\begin{align}
\uparrow _\text{D}\ \leftrightarrow\ (d_{\delta }\! &=+1) \leftrightarrow\ V_\text{D}
&\downarrow _\text{D}\ \leftrightarrow\ (d_{\delta }\! &=-1) \leftrightarrow\ H_\text{D}\notag\\ 
\uparrow _\text{G}\ \leftrightarrow\ (g_{\gamma }\! &=+1) \leftrightarrow\ V_\text{G}
&\downarrow _\text{G}\ \leftrightarrow\ (g_{\gamma }\! &=-1) \leftrightarrow\ H_\text{G}\ ,\label{ksmgnsg}
\end{align} 
a unique numerical notation can be applied to the pairs of results $d_\delta $\,,\bz{\,}$g_\gamma $ of spin- or polarization\bz{-}measurements of the two particles with analyzer settings $\delta $ and $\gamma $\,. 

If both detectors are set to the same angle ($\delta =\gamma $), then 
\begin{align}
d_{\gamma }=\pklammerm g_{\gamma }\ ,\label{isjghsdgh}
\end{align} 
where the sign in brackets is for anticorrelated systems like \eqref{mjsmngnsdc} or \eqref{msnmgnsdgb}, and the sign without brackets for correlated systems like \eqref{mjsmngnsdd} or \eqref{msnmgnsdga}. 

The expectation values $\langle d_{\delta }\cdot g_{\gamma }\rangle $ of the products $d_{\delta }\cdot g_{\gamma }$ are called correlation functions. They are related to the probabilities $P$ of the various pairs of results due to  
\begin{align}
\hspace{-1em}\langle d_{\delta }\cdot g_{\gamma }\rangle =\begin{cases}
P(\,\uparrow _\text{D}\,\uparrow _\text{G})+P(\,\downarrow _\text{D}\,\downarrow _\text{G})-P(\,\downarrow _\text{D}\,\uparrow _\text{G})-P(\,\uparrow _\text{D}\,\downarrow _\text{G})\\ 
P(\text{V}_\text{D}\,\text{V}_\text{G})+P(\text{H}_\text{D}\,\text{H}_\text{G})-P(\text{H}_\text{D}\,\text{V}_\text{G})-P(\text{V}_\text{D}\,\text{H}_\text{G})\ .\end{cases}\hspace{-1em}\label{oskjgnjs}
\end{align} 

\section{\texorpdfstring{\;A\,`proof}{A 'proof} by contradiction'}
Within mathematics, a `proof by contradiction' runs like this: First some basic assumptions A1,\bz{\,}A2,\bz{\,}A3,\bz{\,}\dots\ are made. Then in several steps of derivation D1,\bz{\,}D2,\bz{\,}D3,\bz{\,}\dots\ conclusions are derived from the basic assumptions, and eventually a contradiction is pointed out. The contradiction proves, that at least one of the basic assumptions A1,\bz{\,}A2,\bz{\,}A3,\bz{\,}\dots\ is wrong.

The method of `proof by contradiction' can be transfered to physics, with the contradiction being a contradiction between one of the derived expressions D1,\bz{\,}D2,\bz{\,}D3,\bz{\,}\dots\ and some experimental results. As experimental results are true by definition, the contradiction proves that at least one of the basic assumptions A1,\bz{\,}A2,\bz{\,}A3,\bz{\,}\dots\ is wrong. 

There is a further important difference between mathematical and physical `proofs by contradiction': Due to the axiomatic structure of mathematics, it is easy to identify the complete set of basic assumptions in a mathematical proof. In the physical proof, however, implicit basic assumptions may easily slip attention. Hence we can never be absolutely sure that the contradiction is really caused by one or several of the explicit assumptions but not by some unnoticed implicit assumption. As Einstein phrased it: \al Inasmuch as mathematical theorems refer to reality, they are not certain, and inasmuch as they are certain, they do not refer to reality.\arp\!\cite{Einstein:GeometrieErfahrung}

Bell\!\cite{Bell:Ungleichung} derived his inequality from two explicit basic assumptions:\linebreak\vspace*{-1\baselineskip}  
\vspace{.5ex}\begin{changemargin}{0em}{4.3em}\hspace{-3.34em}\hypertarget{A1Bell}{\textbf{A1}$_\text{Bell}$} The (assumed incomplete) textbook quantum theory can be completed by hidden variables, which uniquely determine the results of single measurements.\vspace{.5ex}\\ 
\hspace*{-3.34em}\hypertarget{A2Bell}{\textbf{A2}$_\text{Bell}$} The measurement result $g_{\gamma }$ is not affected by the apparatus setup nor the result of the measurement of $d_{\delta }$\,, and vice versa, if these measurements are space\bz{-}like separated.
\end{changemargin}
Only many years later Bell became aware\!\cite{Bell:Beables} that he in addition had implicitly made this assumption: 
\vspace{.5ex}\begin{changemargin}{0em}{4.3em}\hspace{-3.34em}\hypertarget{A3Bell}{\textbf{A3}$_\text{Bell}$} The settings of the measurement devices, and the actual outcomes of the measurements, at two (possibly space\bz{-}like separated) locations are not all four determined by a common cause in their common past lightcone (no \al super\bz{-}determinism\arp ). 
\vspace{.5ex}\end{changemargin} 
This assumption may seem self\bz{-}evident, but it is not. The feasibility and advantages of superdeterminism have been evaluated to appreciable detail by Palmer\!\cite{Palmer:superdeterminism}. 

Cramer\!\cite{Cramer:absorbtheor} noted that Bell's reasoning was in addition based onto this implicit assumption: 
\vspace{.5ex}\begin{changemargin}{0em}{4.3em}\hspace*{-3.34em}\hypertarget{A4Bell}{\textbf{A4}$_\text{Bell}$} The future outcome of a measurement does not influence the prior apparatus settings (no \al retrocausation\arp ).
\vspace{.5ex}\end{changemargin}

De\,la\,Pe\~{n}a, Cetto, and Brody\!\cite{Pena:Bell} (and later Nieuwenhuizen\!\cite{Nieuwenhuizen:Bellwrong}, who was not aware of the prior work of de\,la\,Pe\~{n}a et.\,al.) pointed out that Bell furthermore made this implicit assumption: 
\vspace{.5ex}\begin{changemargin}{0em}{4.3em}\hspace{-3.34em}\hypertarget{A1BellS}{\textbf{A1}$'_\text{Bell}$} There exists a common distribution $\rho (\lambda )$ of the sets of hidden variables $\lambda $ of measurements with different apparatus settings, even though these measurements cannot be carried out simultaneously.
\vspace{.5ex}\end{changemargin} 
This is of course not really an independent basic assumption, but a closer specification of \hyperlink{A1Bell}{A1}$_\text{Bell}$. For that reason I name it \hyperlink{A1BellS}{A1$'$}$_\text{Bell}$\,. As Bell derived his inequality not within the framework of quantum theory, but --- assumption \hyperlink{A1Bell}{A1}$_\text{Bell}$ --- within a hidden\bz{-}variables theory (which is a classical theory!), \hyperlink{A1BellS}{A1$'$}$_\text{Bell}$ seems in this context a quite sensible and plausible assumption. Still de\,la\,Pe\~{n}a et.\,al.\ are of course right when claiming that Bell's derivation is of no relevance with regards to hidden\bz{-}variables theories which are not constructed according to \hyperlink{A1BellS}{A1$'$}$_\text{Bell}$\,. I will not pursue this question any further, because the issue became obsolete anyway due to the work of Peres, discussed below. 

Bell implemented the hidden variables $\lambda $ assumed in \hyperlink{A1Bell}{A1}$_\text{Bell}$ and \hyperlink{A1BellS}{A1$'$}$_\text{Bell}$ as continuous normalized parameters with density $\rho (\lambda )$\,:  
\begin{align}
\int\!\dif\!\lambda\:\rho (\lambda )=1\notag 
\end{align} 
Thereby the results \eqref{ksmgnsg} became functions of $\lambda $ 
\begin{align}
d_{\delta }\equiv d_{\delta }(\lambda )=\pm 1\quad ,\quad g_{\gamma }\equiv g_{\gamma }(\lambda )=\pm 1\ ,\notag 
\end{align} 
and the correlation functions \eqref{oskjgnjs} became 
\begin{align}
\langle d_{\delta }\cdot g_{\gamma }\rangle =\int\!\dif\!\lambda\:\rho\, d_{\delta }\, g_{\gamma }\ .\notag 
\end{align} 
Making use of the strict correlation (anticorrelation) 
\begin{align}
d_{\gamma }\stackrel{\eqref{isjghsdgh}}{=}\pklammerm g_{\gamma }\ ,\notag 
\end{align} 
with the sign in brackets for anticorrelated systems, and the sign without brackets for correlated systems, Bell now computed 
\begin{align}
&\Big|\langle d_{\delta }\cdot g_{\gamma }\rangle -\langle d_{\delta }\cdot g_{\gamma '}\rangle\Big| 
=\Big|\int\!\dif\!\lambda\:\rho\,\Big( d_{\delta }\, g_{\gamma }\mklammerp\underbrace{d_{\gamma }\, g_{\gamma }}_{\scalebox{.8}{\pklammerm}1}d_{\delta }\, g_{\gamma '}\Big)\Big|\leq\notag\displaybreak[1]\\ 
&\leq\int\!\dif\!\lambda\:\rho\,\underbrace{\Big|\, d_{\delta }\, g_{\gamma }\Big|}_{1}\,\Big( \underbrace{1\mklammerp d_{\gamma }\, g_{\gamma '}}_{\geq 0}\Big) =1\mklammerp\langle d_{\gamma }\cdot g_{\gamma '}\rangle\ ,\label{osdfkgjg}
\end{align} 
and thereby derived the inequality 
\vspace{1ex}\begin{changemargin}{0em}{4.3em}\hspace{-3.34em}\hypertarget{D1Bell}{\textbf{D1}$_\text{Bell}$}\hspace{2.5em}$\Big|\langle d_{\delta }\cdot g_{\gamma }\rangle -\langle d_{\delta }\cdot g_{\gamma '}\rangle\Big|\:\pklammerm\langle d_{\gamma }\cdot g_{\gamma '}\rangle\leq 1\ .$
\vspace{1ex}\end{changemargin}
In \!\cite[eq.\,(15)]{Bell:Ungleichung} only a $-$ sign shows up instead of $\pklammerm $\,, because Bell considered exclusively the anticorrelated system \eqref{mjsmngnsdc}. When Bell derived this inequality, there existed not yet an experimental test, and hence no contradiction with experiment. But Bell noted that \hyperlink{D1Bell}{D1}$_\text{Bell}$ is not compatible with the predictions of quantum theory. Inserting the quantum\bz{-}theoretical correlation function 
\begin{align}
\langle d_{\delta }\cdot g_{\gamma }\rangle =\pklammerm\cos (\gamma -\delta )\notag 
\end{align} 
of a system of two (anti)correlated \mbox{spin\,\nicefrac{1}{2}} particles into \hyperlink{D1Bell}{D1}$_\text{Bell}$, one gets for example 
\begin{align}
&\text{with }\delta =0\,,\,\gamma =\pi /2\,,\,\gamma '=3\pi /4\,:\notag\\ 
&\Big|\,\pklammerm\cos (\gamma -\delta )\mklammerp\cos (\gamma '-\delta )\,\Big| +\cos (\gamma '-\gamma )=\notag\displaybreak[1]\\  
&=\Big|\,\pklammerm 0\;\pklammerm\!\sqrt{1/2}\,\Big| +\sqrt{1/2}=\sqrt{2}\nleq 1\ .\notag 
\end{align} 
Thus Bell arrived at this conclusion: 
\vspace{.5ex}\begin{changemargin}{0em}{4.3em}\hspace{-3.34em}\hypertarget{C1Bell}{\textbf{C1}$_\text{Bell}$}\hspace{.5em}{\small{}IF} \textbf{\{}\hyperlink{A3Bell}{A3}$_\text{Bell}$ and \hyperlink{A4Bell}{A4}$_\text{Bell}$ are correct,\textbf{\}} {\small{}THEN} \textbf{\{}a theory which assumes both \hyperlink{A1Bell}{A1}$_\text{Bell}$ (with specification \hyperlink{A1BellS}{A1$'$}$_\text{Bell}$) and \hyperlink{A2Bell}{A2}$_\text{Bell}$ can not reproduce all statistical predictions of quantum theory.\textbf{\}}
\vspace{.5ex}\end{changemargin}
Actually Bell did not mention \hyperlink{A3Bell}{A3}$_\text{Bell}$, \hyperlink{A4Bell}{A4}$_\text{Bell}$, and \hyperlink{A1BellS}{A1$'$}$_\text{Bell}$ in his publication\!\cite{Bell:Ungleichung}. Thus these assumptions seemed self\bz{-}evident to him, or he was not aware of them. 

While the inequality \hyperlink{D1Bell}{D1}$_\text{Bell}$ is mathematically perfectly correct, it is not at all convenient for experimental tests. The problem is: To measure the first correlation function, the analyzer of particle G must be set to $\gamma $\,. And to measure the third correlation function, the analyzer of particle D must be set to $\gamma $. A small difference between the two alleged $\gamma $ settings of the two analyzers can hardly be avoided, spoiling the accuracy of the result. Therefore Clauser, Horne, Shimony, and Holt\!\cite{Clauser:bell} derived a modified inequality, which avoids that problem. 

They introduced an additional angle $\delta '$, and split the space of the hidden variables $\lambda $ into regions $\Gamma _+$ and $\Gamma _-$, defined by 
\begin{align}
\Gamma _+\,:\quad d_{\delta '}(\lambda )&=+d_{\gamma }(\lambda )=\pklammerm g_{\gamma }(\lambda )\notag\\ 
\Gamma _-\,:\quad d_{\delta '}(\lambda )&=-d_{\gamma }(\lambda )=\mklammerp g_{\gamma }(\lambda )\ ,\notag 
\end{align} 
where as usual the signs in brackets apply to anticorrelated systems, and the signs without brackets apply to correlated systems. With this notation, 
\begin{align}
&\langle d_{\delta '}\cdot g_{\gamma }\rangle 
=\int\!\dif\!\lambda\:\rho\, d_{\delta '}\, g_{\gamma }
=\int _{\Gamma _+}\hspace{-.6em}\dif\!\lambda\:\rho\,\underbrace{d_{\delta '}\, g_{\gamma }}_{\scalebox{.8}{\pklammerm}1}
+\int _{\Gamma _-}\hspace{-.6em}\dif\!\lambda\:\rho\,\underbrace{d_{\delta '}\, g_{\gamma }}_{\scalebox{.8}{\mklammerp}1}=\notag\\ 
&=\pklammerm\!\int\!\dif\!\lambda\:\rho\,\mklammerp 2\int _{\Gamma _-}\hspace{-.6em}\dif\!\lambda\:\rho\, 
=\pklammerm 1\mklammerp 2\int _{\Gamma _-}\hspace{-.6em}\dif\!\lambda\:\rho\ ,\label{kjsdnmgnsg}\displaybreak[1]\\ 
&\langle d_{\gamma }\cdot g_{\gamma '}\rangle =\int\!\dif\!\lambda\:\rho\, d_{\gamma }\, g_{\gamma '}
=\int_{\Gamma _+}\!\dif\!\lambda\:\rho\, d_{\delta '}\, g_{\gamma '}
-\int_{\Gamma _-}\!\dif\!\lambda\:\rho\, d_{\delta '}\, g_{\gamma '}=\notag\\ 
&=\int\!\dif\!\lambda\:\rho\, d_{\delta '}\, g_{\gamma '}
-2\int_{\Gamma _-}\!\dif\!\lambda\:\rho\, d_{\delta '}\, g_{\gamma '}
\geq\langle d_{\delta '}\cdot g_{\gamma '}\rangle -2\int_{\Gamma _-}\!\dif\!\lambda\:\rho
\stackrel{\eqref{kjsdnmgnsg}}{=}\notag\\ 
&=\langle d_{\delta '}\cdot g_{\gamma '}\rangle\pklammerm\langle d_{\delta '}\cdot g_{\gamma }\rangle -1\ .\label{jmksnmgnsg} 
\end{align} 
Inserting \eqref{jmksnmgnsg} into \hyperlink{D1Bell}{\textbf{D1}}$_\text{Bell}$ gives 
\vspace{.5ex}\begin{changemargin}{0em}{4.3em}\hspace{-3.34em}\hypertarget{D2Bell}{\textbf{D2}$_\text{Bell}$}\hspace{1.5em}$\Big|\langle d_{\delta }\cdot g_{\gamma }\rangle -\langle d_{\delta }\cdot g_{\gamma '}\rangle\Big|\:\pklammerm\langle d_{\delta '}\cdot g_{\gamma '}\rangle +\langle d_{\delta '}\cdot g_{\gamma }\rangle \leq 1\pklammerm 1\ .$
\vspace{.5ex}\end{changemargin}
In \!\cite[eq.\,(1a)]{Clauser:bell} $+$ signs shows up instead of $\pklammerm $\,, because CHSH considered exclusively the correlated system \eqref{msnmgnsdga}, but no anticorrelated system. The different signs for correlated versus anticorrelated systems can be avoided by adding 
\begin{align}
+2\,\langle d_{\delta '}\cdot g_{\gamma '}\rangle\leq +2\notag 
\end{align}
to the inequality for the anticorrelated case. Thereby one gets the inequality 
\vspace{.5ex}\begin{changemargin}{0em}{4.3em}\hspace{-3.34em}\hypertarget{D3Bell}{\textbf{D3}$_\text{Bell}$}\hspace{1.5em}$\Big|\langle d_{\delta }\cdot g_{\gamma }\rangle -\langle d_{\delta }\cdot g_{\gamma '}\rangle\Big|\: +\langle d_{\delta '}\cdot g_{\gamma '}\rangle +\langle d_{\delta '}\cdot g_{\gamma }\rangle \leq 2\ ,$
\vspace{.5ex}\end{changemargin}
which is valid for both anticorrelated and correlated systems. Note that I continue to use the in\-{}dex$_\text{Bell}$\,, because CHSH derived this inequality from Bell's basic assumptions. 
The inequality \hyperlink{D3Bell}{D3}$_\text{Bell}$ has meanwhile been checked by numerous experiments, \ggeg\cite{Aspect:correlation81,Aspect:correlation82a,Aspect:correlation82b,Rowe:BellTest,Kwia:typeIIdc,Weihs:BellExperi,Hensen:Belltest,Giustina:Belltest,Shalm:Belltest,WenjaminRosenfeld:BellTest}. The experimental results are (within measurement errors) consistent with quantum theory, but violate \hyperlink{D3Bell}{D3}$_\text{Bell}$ significantly. Thus there is a 
\vspace{.5ex}\begin{changemargin}{0em}{8.6em}\hspace{-7.8em}\textbf{Contradiction: }\hyperlink{D3Bell}{D3}$_\text{Bell}$ is not compatible with numerous experimental results.
\vspace{.5ex}\end{changemargin}
This contradiction implies the conclusion that at least one of Bell's five basic assumptions is wrong. Note that the `proof by contradiction' does not indicate which of the five basic assumptions is\bz{/}are wrong. But we may group the basic assumptions into those which seem to be most likely correct, \ggie \hyperlink{A3Bell}{A3}$_\text{Bell}$\,,\bz{\,}\hyperlink{A4Bell}{A4}$_\text{Bell}$, and a second group of assumptions which seem \al problematic\arp , \ggie \hyperlink{A1Bell}{A1}$_\text{Bell}$ (with specification \hyperlink{A1BellS}{A1$'$}$_\text{Bell}$)\,,\bz{\,}\hyperlink{A2Bell}{A2}$_\text{Bell}$\,. Thereby the conclusion may be formulated like this: 
\vspace{.5ex}\begin{changemargin}{0em}{4.3em}\hspace{-3.3em}\hypertarget{C2Bell}{\textbf{C2}$_\text{Bell}$}\hspace{.5em}{\small IF} \textbf{\{}\hyperlink{A3Bell}{A3}$_\text{Bell}$ and \hyperlink{A4Bell}{A4}$_\text{Bell}$ are correct,\textbf{\}} {\small THEN} \textbf{\{}\textbf{\{}quantum theory can \emph{not} be completed by hidden variables (as specified in \hyperlink{A1BellS}{A1$'$}$_\text{Bell}$) which uniquely determine the results of single measurements,\textbf{\}} {\small OR} \textbf{\{}the measurement result $g_{\gamma }$ \emph{is} affected by the apparatus setup and the result of the measurement of $d_{\delta }$\,, and vice versa, even if these measurements are space\bz{-}like separated.\textbf{\}}\textbf{\}} 
\vspace{.5ex}\end{changemargin}
Note that the {\small OR} is an inclusive {\small OR}, but not an exclusive {\small EXOR}. If \hyperlink{A3Bell}{A3}$_\text{Bell}$ and \hyperlink{A4Bell}{A4}$_\text{Bell}$ are correct, then the experimental violation of \hyperlink{D3Bell}{D3}$_\text{Bell}$ may imply that quantum theory can not be completed by hidden variables of the specified type, or that even space\bz{-}like separated measurements of $d_{\delta }$ and $g_{\gamma }$ are not mutually independent, or both. 

\section{The derivations of Bell's inequality due to Wigner and Peres}
Wigner\!\cite{Wigner:HiddenVariab} derived an inequality of the Bell's type from quite abstract set\bz{-}theoretical mathematical considerations. While that derivation is elegant, the application of the results to physical systems is somewhat delicate, and caused doubts and objections\!\cite{Hess:Wigner}. To avoid these problems, I will present in the sequel Wigner's derivation based onto the basic assumptions as proposed by Peres. 

Peres\!\cite{Peres:BellTheorem} based the derivation of Bell's inequality onto only one explicit basic assumption: 
\vspace{.5ex}\begin{changemargin}{0em}{4.3em}\hspace{-3.7em}\hypertarget{A1Peres}{\textbf{A1}$_\text{Peres}$} If the result of a measurement can be predicted with probability $P=1$\,, then that result is --- even if the measurement is not actually performed --- as real as the result of an actually performed measurement.
\vspace{.5ex}\end{changemargin}
Peres did not use exactly this wording, but his assumption is equivalent to this formulation. Notions like \al real\ar  or \al exist\ar  are hard to define, hence a source of much confusion in interpretations of quantum phenomena. It is an important advantage of assumption \hyperlink{A1Peres}{A1}$_\text{Peres}$ that it circumvents this problem due to the formulation \al as real as \dots\arp . \hyperlink{A1Peres}{A1}$_\text{Peres}$ says that the results of not performed measurements --- provided they can be predicted with probability unity --- share the same status of reality as actually measured results, whatever that status of reality may be. 

Assumption \hyperlink{A1Peres}{A1}$_\text{Peres}$  alludes directly to the argument forwarded by Einstein, Podolski, and Rosen\!\cite{Einstein:EPR}. EPR had insisted that measurement results for arbitrary settings of the apparatus angles $\delta $ and $\gamma $ must be considered \al parts of reality\arp , because the result of any measurement can be predicted with probability $P=1$ by simply measuring the correlated second particle with the same angle setting ($\delta =\gamma $), and making use of the strict (anti)\-cor\-relat\-ion \eqref{isjghsdgh}. This is always possible, even if the measurements are performed space\bz{-}like separated. 

EPR believed that the mere possibility to perform such measurements and predict their results with probability unity was sufficient to make those results \al parts of reality\arp , even if these measurements were not actually performed. I will only later discuss the similarity and the differences of \hyperlink{A1Peres}{A1}$_\text{Peres}$ versus \hyperlink{A1Bell}{A1}$_\text{Bell}$ (including \hyperlink{A1BellS}{A1$'$}$_\text{Bell}$). 

Wigner and Peres definitively did not adopt Bell's \al separability\ar  assumption \hyperlink{A2Bell}{A2}$_\text{Bell}$\,. But like Bell they implicitly assumed \al no superdeterminism\ar  and \al no retrocausation\arp :  
\vspace{.5ex}\begin{changemargin}{0em}{4.3em}\hspace{-3.7em}\hypertarget{A2Peres}{\textbf{A2}$_\text{Peres}$} The settings of the measurement devices, and the actual outcomes of the measurements, at two (possibly space\bz{-}like separated) locations are not all four determined by a common cause in their common past lightcone (no \al super\bz{-}determinism\arp ). \vspace{1ex}\\ 
\hspace*{-3.7em}\hypertarget{A3Peres}{\textbf{A3}$_\text{Peres}$} The future outcome of a measurement does not influence the prior apparatus settings (no \al retrocausation\arp ).
\vspace{.5ex}\end{changemargin}

As before we use the notations $d_{\delta}=\pm 1$ and $g_{\gamma}=\pm 1$ for the results of actually performed measurements. According to \hyperlink{A1Peres}{A1}$_\text{Peres}$\,, the results $d_{\delta '}$,\bz{\,}$d_{\delta ''}$,\bz{\,}$d_{\delta '''}$,\bz{\,}\dots\ with arbitrary angles different from $\delta $, and the results $g_{\gamma '}$,\bz{\,}$g_{\gamma ''}$,\bz{\,}$g_{\gamma '''}$,\bz{\,}\dots\ with arbitrary angles different from $\gamma $, are --- even though the measurements with these angle settings are not actually performed --- in each experimental run as real as the actually measured results $d_{\delta}$ and $g_{\gamma}$\,. 

For Wigner's derivation it is sufficient to conclude from \hyperlink{A1Peres}{A1}$_\text{Peres}$ that in addition to the actually measured results $d_{\delta }$ and $g_{\gamma }$\,, two additional  not actually measured results $d_{\delta '}=\pm 1$ and $d_{\delta ''}=\pm 1$ exist for particle D, and two additional  not actually measured results $g_{\gamma '}=\pm 1$ and $g_{\gamma ''}=\pm 1$ exist for particle G\,. This implies the conclusion 
\vspace{.5ex}\begin{changemargin}{0em}{4.3em}\hspace{-3.8em}\hypertarget{D1Peres}{\textbf{D1}$_\text{Peres}$}\hspace{.5em}In each experimental run not only the doublet \\ 
\hspace*{8em}$(d_{\delta }\,,\,g_{\gamma })$\\ 
of the two actually measured results, but the full sextet \\ 
\hspace*{4.5em}$(d_{\delta }\,,\, d_{\delta '}\,,\,d_{\delta ''}\; ;\; g_{\gamma }\,,\,g_{\gamma '}\,,\,g_{\gamma ''})$\\ 
with two actually measured results, and four additional not measured results, is as real as the doublet of the actually measured results.
\vspace{.5ex}\end{changemargin}
Note that I continue to use the index$_\text{Peres}$\,, because (in my presentation) Wigner's derivation is based onto Peres' basic assumptions. Wigner considered spin projections onto three linearly independent space axes, but the linear independency is of no relevance for his argument. Thus we continue to consider exclusively projections onto the $xy$\bz{-}plane, see fig.\;\ref{fig:Setup}\,. Furthermore Wigner chose for spin measurements of both particles D and G the same three angles $\vartheta _1$\,,\bz{\,}$\vartheta _2$\,,\bz{\,}$\vartheta _3$\,. Making use of the strict (anti)correlation 
\begin{align}
d_{\vartheta _j}\stackrel{\eqref{isjghsdgh}}{=}\pklammerm g_{\vartheta _j}\ ,\label{jnsngnbsd}
\end{align} 
where the sign in brackets is for anticorrelated systems, and the sign without brackets for correlated systems, abbreviating $\pm 1$ by $\pm $, and using $?$ as a wildcard, a typical result sextet thus may be written as 
\begin{align}
\Big( d_{\vartheta _1}\,,\, d_{\vartheta _2}\,,\,d_{\vartheta _3}\;\scalebox{1.4}{;}\; g_{\vartheta _1}\,,\,g_{\vartheta _2}\,,\,g_{\vartheta _3}\Big) =\Big(\,\framebox[6mm]{\rule[-.3mm]{0mm}{2.1mm}$-$}\,,\, ?\,,\,\pklammerm\; \scalebox{1.4}{;}\;\mklammerp\,,\, ?\,,\,\framebox[6mm]{\rule[-.3mm]{0mm}{2.1mm}$+$}\,\Big)\ .\notag 
\end{align}
The actually measured results are marked by boxes. In this example, the $\vartheta _1$\bz{-}component of particle D was measured with result $-$, and the $\vartheta _1$\bz{-}component of particle G was concluded from \eqref{jnsngnbsd} as $\mklammerp $. The $\vartheta _2$\bz{-}components of neither particle were measured, and therefore got the wildcards. Still these components are assumed to exist as real as the measured components, and have the values $+$ or $-$\,. The $\vartheta _3$\bz{-}component of particle G was measured with result $+$, and the $\vartheta _3$\bz{-}component of particle D was concluded from \eqref{jnsngnbsd} as $\pklammerm $. 

Next Wigner considered the probabilities of some particular result sextets, \ggie  the probabilities that this particular result sextet will be realized in a single experimental run. I will use the notation \mbox{$(\dots ;\dots )$} with round brackets for the result sextets, and the notation \mbox{$\langle\dots ;\dots\rangle $} with angle brackets for the probabilities of the sextets: 
 \begin{subequations}\begin{align}
&\Big\langle\,\framebox[6mm]{\rule[-.3mm]{0mm}{2.1mm}$+$}\,,\,\pklammerm\,,\,?\;\scalebox{1.4}{;}\;\pklammerm\,,\,\,\framebox[6mm]{\rule[-.3mm]{0mm}{2.1mm}$+$}\,,\, ?\,\Big\rangle =\notag\\ 
&=\Big\langle +\,,\,\pklammerm\,,\,+\;\scalebox{1.4}{;}\;\pklammerm\,,\, +\,,\,\pklammerm\Big\rangle 
+\Big\langle +\,,\,\pklammerm\,,\,-\;\scalebox{1.4}{;}\;\pklammerm\,,\, +\,,\,\mklammerp\Big\rangle\notag\\ 
&\geq\Big\langle +\,,\,\pklammerm\,,\,-\;\scalebox{1.4}{;}\;\pklammerm\,,\, +\,,\,\mklammerp\Big\rangle\label{jksdjngna}  
\end{align}
The inequality holds, because any probability is positive definite. Likewise one gets these probabilities: 
\begin{align}
&\Big\langle\,\framebox[6mm]{\rule[-.3mm]{0mm}{2.1mm}$-$}\,,\, ?\,,\, -\;\scalebox{1.4}{;}\;\mklammerp\,,\, ?\,,\,\framebox[6mm]{\rule[-.3mm]{0mm}{2.1mm}\mklammerp}\,\Big\rangle =\notag\\ 
&=\Big\langle -\,,\,\pklammerm\,,\, -\;\scalebox{1.4}{;}\;\mklammerp\,,\, +\,,\,\mklammerp\Big\rangle 
+\Big\langle -\,,\,\mklammerp\,,\, -\;\scalebox{1.4}{;}\;\mklammerp\,,\, -\,,\,\mklammerp\Big\rangle \notag\\ 
&\geq\Big\langle -\,,\,\pklammerm\,,\, -\;\scalebox{1.4}{;}\;\mklammerp\,,\, +\,,\,\mklammerp\Big\rangle \label{jksdjngnb}\displaybreak[1]\\  
&\Big\langle\, ?\,,\,\pklammerm\,,\,\,\framebox[6mm]{\rule[-.3mm]{0mm}{2.1mm}$-$}\ \scalebox{1.4}{;}\ ?\,,\,\,\framebox[6mm]{\rule[-.3mm]{0mm}{2.1mm}$+$}\,,\,\mklammerp\Big\rangle =\label{jksdjngnc}\\ 
&=\Big\langle +\,,\,\pklammerm\,,\,-\;\scalebox{1.4}{;}\;\pklammerm\,,\, +\,,\,\mklammerp\Big\rangle 
+\Big\langle -\,,\,\pklammerm\,,\, -\;\scalebox{1.4}{;}\;\mklammerp\,,\, +\,,\,\mklammerp\Big\rangle \notag 
\end{align}\end{subequations} 
Inserting \eqref{jksdjngna} and \eqref{jksdjngnb} into \eqref{jksdjngnc}, Wigner got 
\vspace{.5ex}\begin{changemargin}{0em}{4.3em}\hspace{-3.34em}\hypertarget{D2Peres}{\textbf{D2}$_\text{Peres}$}\hspace{2.5em}$\Big\langle\, ?\,,\,\pklammerm\,,\,\,\framebox[6mm]{\rule[-.3mm]{0mm}{2.1mm}$-$}\ \scalebox{1.4}{;}\ ?\,,\,\,\framebox[6mm]{\rule[-.3mm]{0mm}{2.1mm}$+$}\,,\,\mklammerp\Big\rangle \leq $\vspace{.8ex}\\ 
\hspace*{5em}$\leq\Big\langle\,\framebox[6mm]{\rule[-.3mm]{0mm}{2.1mm}$+$}\,,\,\pklammerm\,,\,?\;\scalebox{1.4}{;}\;\pklammerm\,,\,\,\framebox[6mm]{\rule[-.3mm]{0mm}{2.1mm}$+$}\,,\, ?\,\Big\rangle +\vspace{.8ex}\\ 
\hspace*{6em}+\Big\langle\,\framebox[6mm]{\rule[-.3mm]{0mm}{2.1mm}$-$}\,,\, ?\,,\, -\;\scalebox{1.4}{;}\;\mklammerp\,,\, ?\,,\,\framebox[6mm]{\rule[-.3mm]{0mm}{2.1mm}\mklammerp}\,\Big\rangle $\ . 
\vspace{.5ex}\end{changemargin}
When Wigner derived this inequality in 1970, there existed not yet any experimental result, to which the inequality could be compared. But Wigner noted that \hyperlink{D2Peres}{D2}$_\text{Peres}$ is not compatible with the predictions of quantum theory. He considered two anticorrelated \mbox{spin\,\nicefrac{1}{2}} particles in the singlet state \eqref{mjsmngnsdc}. The probability, that a Stern\bz{-}Gerlach magnet set to $\vartheta _1$ will measure the first particle as $+$ is simply 1/2\,. The probability that the anticorrelated partner particle will be measured as well as $+$ with it's Stern\bz{-}Gerlach magnet set to $\vartheta _2$ is $\sin ^2(\vartheta _2/2-\vartheta _1/2)$ according to quantum theory. Inserting these quantum\bz{-}theoretical expectation values into \hyperlink{D2Peres}{D2}$_\text{Peres}$\,, one gets for the anticorrelated \mbox{spin\,\nicefrac{1}{2}} particles: 
\begin{align}
&\frac{1}{2}\,\cos ^2\Big(\frac{\vartheta _2-\vartheta _3}{2}\Big)\stackrel{\displaystyle\boldsymbol{?}}{\leq}
\frac{1}{2}\,\sin ^2\Big(\frac{\vartheta _2-\vartheta _1}{2}\Big) +\frac{1}{2}\,\cos ^2\Big(\frac{\vartheta _3-\vartheta _1}{2}\Big)\label{jndnjhfgnbsddg} 
\end{align}
This inequality is violated with many angle settings. For example, with $\vartheta _1=0^\circ\, $ and $\vartheta _3=90^\circ\, $, \eqref{jndnjhfgnbsddg} is violated for all $0^\circ <\vartheta _2<90^\circ\, $, with a maximum violation at $\vartheta _2=45^\circ\, $. Thus we arrive at this conclusion: 
\vspace{.5ex}\begin{changemargin}{0em}{4.3em}\hspace{-3.8em}\hypertarget{C1Peres}{\textbf{C1}$_\text{Peres}$}\hspace{.5em}{\small{}IF} \textbf{\{}\hyperlink{A2Peres}{A2}$_\text{Peres}$ and \hyperlink{A3Bell}{A3}$_\text{Peres}$ are correct,\textbf{\}} {\small{}THEN} \textbf{\{}any theory, which assumes that the results of not performed measurements --- provided these results can be predicted with probability unity --- are as real as the results of actually performed measurements, can impossibly reproduce all statistical predictions of quantum theory.\textbf{\}}
\vspace{.5ex}\end{changemargin}
Wigner did not mention \hyperlink{A2Peres}{A2}$_\text{Peres}$ and \hyperlink{A3Peres}{A3}$_\text{Peres}$ in his publication\!\cite{Wigner:HiddenVariab}. Thus these assumptions seemed self\bz{-}evident to him, or he was not aware of them. 

Peres\!\cite{Peres:BellTheorem} generalized Wigner's derivation to four different angle settings, such that the analyzer of one particle does not need to be set to exactly the same angle as the analyzer of the other particle, thereby making the inequality better amenable to experimental tests. For Peres' derivation it is sufficient to conclude from \hyperlink{A1Peres}{A1}$_\text{Peres}$ that just one not actually measured result $d_{\delta '}=\pm 1$ with $\delta '\neq\delta $ and one not actually measured result $g_{\gamma '}=\pm 1$ with $\gamma '\neq\gamma $ exist and share the same status of reality as the actually measured results $d_{\delta }$ and $g_{\gamma }$\,. This implies the conclusion 
\vspace{.5ex}\begin{changemargin}{0em}{4.3em}\hspace{-3.8em}\hypertarget{D3Peres}{\textbf{D3}$_\text{Peres}$}\hspace{.5em}In each experimental run not only the doublet \\ 
\hspace*{8em}$(d_{\delta }\,,\,g_{\gamma })$\\ 
of the two actually measured results, but the full quartet \\ 
\hspace*{6em}$(d_{\delta }\,,\,g_{\gamma }\,,\,d_{\delta '}\,,\,g_{\gamma '})$\\ 
with two actually measured results, and two additional not measured results, is as real as the doublet of the actually measured results.
\vspace{.5ex}\end{changemargin}
In each experimental run, the result quartet must be one of the 16 different quartets displayed in the 16 columns of table\;\ref{tab:quartets} on the following page. 
\begin{table}[htb]
\centering\small\begin{tabular}{c|r@{\hspace*{.5em}}r@{\hspace*{.5em}}r@{\hspace*{.5em}}r@{\hspace*{.5em}}r@{\hspace*{.5em}}r@{\hspace*{.5em}}r@{\hspace*{.5em}}r@{\hspace*{.5em}}r@{\hspace*{.5em}}r@{\hspace*{.5em}}r@{\hspace*{.5em}}r@{\hspace*{.5em}}r@{\hspace*{.5em}}r@{\hspace*{.5em}}r@{\hspace*{.5em}}r}
 &{\scriptsize 1}\hspace*{.4em}&{\scriptsize 2}\hspace*{.4em}&{\scriptsize 3}\hspace*{.4em}&{\scriptsize 4}\hspace*{.4em}&{\scriptsize 5}\hspace*{.4em}&{\scriptsize 6}\hspace*{.4em}&{\scriptsize 7}\hspace*{.4em}&{\scriptsize 8}\hspace*{.4em}&{\scriptsize 9}\hspace*{.4em}&{\scriptsize 10}\hspace*{.3em}&{\scriptsize 11}\hspace*{.3em}&{\scriptsize 12}\hspace*{.3em}&{\scriptsize 13}\hspace*{.3em}&{\scriptsize 14}\hspace*{.3em}&{\scriptsize 15}\hspace*{.3em}&{\scriptsize 16}\hspace*{.3em}\\ 
\hline 
\hspace*{-.5em}{\normalsize $d_{\delta }$}\hspace*{-.3em}&\hspace*{-.3em}$+1$&$+1$&$+1$&$+1$&$+1$&$+1$&$+1$&$+1$&$-1$&$-1$&$-1$&$-1$&$-1$&$-1$&$-1$&$-1$\\ 
\hspace*{-.5em}{\normalsize $g_{\gamma }$}\hspace*{-.3em}&\hspace*{-.3em}$+1$&$+1$&$+1$&$+1$&$-1$&$-1$&$-1$&$-1$&$+1$&$+1$&$+1$&$+1$&$-1$&$-1$&$-1$&$-1$\\ 
\hspace*{-.5em}{\normalsize $d_{\delta '}$}\hspace*{-.3em}&\hspace*{-.3em}$+1$&$+1$&$-1$&$-1$&$+1$&$+1$&$-1$&$-1$&$+1$&$+1$&$-1$&$-1$&$+1$&$+1$&$-1$&$-1$\\ 
\hspace*{-.5em}{\normalsize $g_{\gamma '}$}\hspace*{-.3em}&\hspace*{-.3em}$+\raisebox{0mm}[0mm][2mm]{1}$&$-1$&$+1$&$-1$&$+1$&$-1$&$+1$&$-1$&$+1$&$-1$&$+1$&$-1$&$+1$&$-1$&$+1$&$-1$\\ 
\hline 
\hspace*{-.5em}$S$\hspace*{-.3em}&\hspace*{-.3em}$+2$&$+2$&$+2$&$-2$&$-2$&$-2$&$+2$&$-2$&$-2$&$+2$&$-2$&$-2$&$-2$&$+2$&$+2$&$+2$ 
\end{tabular}\caption{The 16 quartets $(d_{\delta }\,,\,g_{\gamma }\,,\,d_{\delta '}\,,\,g_{\gamma '})$}\label{tab:quartets}\end{table}

In the bottom line of table\;\ref{tab:quartets}\,, the values of 
\begin{align}
S\equiv d_{\delta }\cdot g_{\gamma }+d_{\delta }\cdot g_{\gamma '}+d_{\delta '}\cdot g_{\gamma }-d_{\delta '}\cdot g_{\gamma '}\label{nsngnbsdg}  
\end{align} 
are displayed for each quartet. As table\;\ref{tab:quartets} is exhaustive (there is no quartet, which is not displayed in this table), we know for sure that 
\begin{align}
S=+2\quad\text{or}\quad S=-2  
\end{align} 
in any single measurement. Consequently we arrive at the following conclusion for the expectation value of $S$, \ggie  for the mean value of $S$ in a huge ensemble of experimental runs: 
\vspace{.5ex}\begin{changemargin}{0em}{4.3em}\hspace{-3.8em}\hypertarget{D4Peres}{\textbf{D4}$_\text{Peres}$} \vspace{-1.8\baselineskip}\begin{align}
-2&\leq\langle\, S\,\rangle\leq +2\notag\\ 
\langle\, S\,\rangle &\equiv \langle d_{\delta }\cdot g_{\gamma }\rangle +\langle d_{\delta }\cdot g_{\gamma '}\rangle +\langle d_{\delta '}\cdot g_{\gamma }\rangle -\langle d_{\delta '}\cdot g_{\gamma '}\rangle\notag
\end{align}
\vspace{-4ex}\end{changemargin}

The inequality \hyperlink{D4Peres}{D4}$_\text{Peres}$ is incompatible with the predictions of quantum theory. 
More important, \hyperlink{D4Peres}{D4}$_\text{Peres}$ has been significantly disproved by experiments, \ggeg\cite{Aspect:correlation81,Aspect:correlation82a,Aspect:correlation82b,Rowe:BellTest,Kwia:typeIIdc,Weihs:BellExperi,Hensen:Belltest,Giustina:Belltest,Shalm:Belltest,WenjaminRosenfeld:BellTest}. Thus there is a 
\vspace{.5ex}\begin{changemargin}{0em}{8.6em}\hspace{-7.8em}\textbf{Contradiction: }\hyperlink{D4Peres}{D4}$_\text{Peres}$ is not compatible with numerous experimental results.
\vspace{.5ex}\end{changemargin}
This contradiction implies the conclusion, that at least one of Peres' three basic assumptions is wrong, provided that no unknown problematic implicit assumption has slipped our attention. Again I give the conclusion a conditional formulation, assuming that \hyperlink{A2Peres}{A2}$_\text{Peres}$ and \hyperlink{A3Peres}{A3}$_\text{Peres}$ most likely are correct: 
\vspace{.5ex}\begin{changemargin}{0em}{4.3em}\hspace{-3.6em}\hypertarget{C2Peres}{\textbf{C2}$_\text{Peres}$} {\small IF} \textbf{\{}\hyperlink{A2Peres}{A2}$_\text{Peres}$ and \hyperlink{A3Peres}{A3}$_\text{Peres}$ are correct,\textbf{\}} {\small THEN} \textbf{\{}the results of not performed measurements do not share the same status of reality as the results of actually performed measurements, not even if the results of the unperformed measurements can be predicted with probability $P\! =\! 1$\,.\textbf{\}}\linebreak\vspace*{-1\baselineskip} 
\vspace{.5ex}\end{changemargin}
The title \al unperformed experiments have no results\ar  of Peres' publication\!\cite{Peres:BellTheorem} reflects this conclusion. 

Besides others, Khrennikov\!\cite{Khrennikov:BellBoole} and Hess, de\,Raedt, and Michiel\-sen\!\cite{Hess:Wigner} raised objections against the argumentation of Wigner and Peres. Khrennikov identified \al probabilistic incompatibilities\arp , because in the derivations of the inequalities \al statistical data from a number of experiments performed under different experimental contexts\ar  are mixed. Quite similar, Hess\bz{\,}et.\,al.\ noted \al that Wigner’s assumptions about the existence of certain joint probabilities are incorrect\arp , and conclude: \al\dots we believe to have shown beyond any reasonable doubt that Wigner derived [\dots ] his Bell type inequality from set theoretically unjustified assumptions about the existence of joint probabilities.\ar  

These objections point out the fact, that the various components of the result sextets and result quartets of Wigner and Peres can only be realized in mutually excluding experimental contexts. Hence the basic assumption \hyperlink{A1Peres}{A1}$_\text{Peres}$ is highly problematic from a quantum theoretical point of view, and actually turned out wrong in the end. 

Still the cited objections are not valid, because the authors did not notice that Wigner and Peres presented their arguments in the particular form of `proofs by contradiction'. The wrong basic assumption \hyperlink{A1Peres}{A1}$_\text{Peres}$ (which implies that \al joint probabilities exist\ar  for arbitrary spin or polarization components, even though the actual measurements of these components would require incompatible experimental settings) does not at all flaw the validity of the proof. 
A `proof by contradiction' can only be flawed by an error in the derived expressions D1\,,\bz{\,}D2\,,\bz{\,}\dots\ One or several wrong basic assumptions A1\,,\bz{\,}A2\,,\bz{\,}\dots , however, are not a fault but --- by construction --- a necessary integral feature of any `proof by contradiction'. Therefore, being a conclusion from a `proof by contradiction', \hyperlink{C2Peres}{C2}$_\text{Peres}$ is perfectly valid and not at all flawed by the wrong basic assumption \hyperlink{A1Peres}{A1}$_\text{Peres}$\,. 

Another objection, which is turning up again and again since decades, e.\,g.\!\cite{Pena:Bell,Schuermann:Bell}, is pointing out that the experimental tests do not exactly match \hyperlink{D4Peres}{D4}$_\text{Peres}$\,. That inequality has been derived under the premise, that only $d_{\delta }$ and $g_{\gamma }$ are measured results, while $d_{\delta '}$ and $g_{\gamma '}$ are not actually measured results. But of course an experimental test of \hyperlink{D4Peres}{D4}$_\text{Peres}$ is possible only, if results with all four analyzer angle settings are actually measured and inserted into that inequality. 

This objection, however, amounts to assume that not actually measured results are as real as actually measured results (assumption \hyperlink{A1Peres}{A1}$_\text{Peres}$), {\small AND} to to assume at the same time that the not measured results, whose values --- if they would be measured --- could be predicted with probability $P\! =\! 1$, would differ from those predicted values if they are not actually measured, even though they are in either case as real as actually measured results. 

With such intricate assumptions about really existing not measured results, however, which magically change whenever they are actually measured, any arbitrary nonsense could be asserted. These considerations become sensible only, if we assume that the not measured, but still really existing results are identical to the results we get when those measurements are actually performed. Hence that objection against the experimental checks of \hyperlink{D4Peres}{D4}$_\text{Peres}$ is not valid. 

In the sequel I will outline four further conclusions \hyperlink{C3Peres}{C3}$_\text{Peres}$\,,\bz{\,}\hyperlink{C4Peres}{C4}$_\text{Peres}$\,,\bz{\,}\hyperlink{C5Peres}{C5}$_\text{Peres}$\,,\bz{\,}\hyperlink{C6Peres}{C6}$_\text{Peres}$\,. These are not independent conclusions, instead they are implicitly enclosed in \hyperlink{C2Peres}{C2}$_\text{Peres}$\,. Because of their importance for our understanding of the world in which we are living, however, it seems worthwhile to make them explicit. 

The fact, that the results of not performed measurements are not as real as the results of actually performed measurements (\hyperlink{C2Peres}{C2}$_\text{Peres}$), immediately implies that the results of measurements do not yet exist before the measurements. Instead the results of actually performed measurements are \emph{created} only in the very moment of measurement, while no results of unperformed measurements are created at any time. 
\vspace{.5ex}\begin{changemargin}{0em}{4.3em}\hspace{-3.6em}\hypertarget{C3Peres}{\textbf{C3}$_\text{Peres}$} {\small IF} \textbf{\{}\hyperlink{A2Peres}{A2}$_\text{Peres}$ and \hyperlink{A3Peres}{A3}$_\text{Peres}$ are correct,\textbf{\}} {\small THEN} \textbf{\{}the results of measurements are not determined before the measurements, but are \emph{created} only in the very moment of measurement.\textbf{\}} 
\vspace{.5ex}\end{changemargin}

Bell's basic assumption \hyperlink{A1Bell}{A1}$_\text{Bell}$ says, that the (assumed incomplete) textbook quantum theory can be completed by hidden variables, which uniquely determine the results of single measurements at arbitrary analyzer settings. Whether or not \hyperlink{A1Bell}{A1}$_\text{Bell}$ is impacted by \hyperlink{C2Peres}{C2}$_\text{Peres}$ and \hyperlink{C3Peres}{C3}$_\text{Peres}$ depends on the precise definition of the notion \al hidden variables\arp . Einstein, Podolski, and Rosen\!\cite{Einstein:EPR} had expected that hidden variables should be related to quantum theory like the molecules of a gas are related to classical thermodynamics: Though $19^\text{th}$ century physicists were not able to verify the existence of molecules, they assumed them to be not merely abstract mathematical constructions on the paper of the theoretical physicist, but really to exist \al out there\arp , and cause macroscopic properties of the gas like pressure and temperature. Such \al really existing\ar  hidden variables, which determine the results of quantum measurements, are clearly excluded by \hyperlink{C2Peres}{C2}$_\text{Peres}$ and \hyperlink{C3Peres}{C3}$_\text{Peres}$\,: If the results of measurements are created only in the moment of measurement, then the results can not be encoded in hidden variables which already are existing before the measurements. 
\vspace{.5ex}\begin{changemargin}{0em}{4.3em}\hspace{-3.7em}\hypertarget{C4Peres}{\textbf{C4}$_\text{Peres}$}\ {\small IF} \textbf{\{}\hyperlink{A2Peres}{A2}$_\text{Peres}$ and \hyperlink{A3Peres}{A3}$_\text{Peres}$ are correct,\textbf{\}} {\small THEN} \textbf{\{}hidden variables, which uniquely determine the results of single quantum measurements, do not correspond to anything \al really existing out there\arp .\textbf{\}} 
\vspace{.5ex}\end{changemargin}

In the introduction of his article\!\cite{Bell:Ungleichung}, Bell had mentioned the theory of Bohm\!\cite{Bohm:hidvar} as an example for a hidden\bz{-}variables theory. If \hyperlink{A2Peres}{A2}$_\text{Peres}$ and \hyperlink{A3Peres}{A3}$_\text{Peres}$ are correct, then nothing \al out there\ar  corresponds to Bohm's quantum potential. As Bohm carefully designed his theory to make exactly the same predictions for experimental results as textbook quantum theory, \hyperlink{C4Peres}{C4}$_\text{Peres}$ says that his hidden\bz{-}variables construction is nothing but a pointless mathematical exercise and idle formal complication of textbook quantum theory, if \hyperlink{A2Peres}{A2}$_\text{Peres}$ and \hyperlink{A3Peres}{A3}$_\text{Peres}$ are correct. Bell's basic assumption \hyperlink{A1Bell}{A1}$_\text{Bell}$ is correct (as demonstrated by Bohm's theory), if \al hidden variables\ar  are interpreted as purely formal constructs on the paper of the theorist. But if Bell assumed (I guess he did) that something exists \al out there\arp, that corresponds to hidden variables, then \hyperlink{A1Bell}{A1}$_\text{Bell}$ is definitively wrong, if \hyperlink{A2Peres}{A2}$_\text{Peres}$ and \hyperlink{A3Peres}{A3}$_\text{Peres}$ are correct. 

Experimental tests\!\cite{Weihs:BellExperi,Hensen:Belltest,Giustina:Belltest,Shalm:Belltest,WenjaminRosenfeld:BellTest} have demonstrated that the inequality \hyperlink{D4Peres}{D4}$_\text{Peres}$ is significantly violated even if the settings of the analyzers and the measurements of the two entangled particles are performed space\bz{-}like separated. Combination of these experimental findings with the fact that measurement results are not determined before the very moment of measurement (\hyperlink{C3Peres}{C3}$_\text{Peres}$) implies the conclusion 
\vspace{.5ex}\begin{changemargin}{0em}{4.3em}\hspace{-3.6em}\hypertarget{C5Peres}{\textbf{C5}$_\text{Peres}$}\ {\small IF} \textbf{\{}\hyperlink{A2Peres}{A2}$_\text{Peres}$ and \hyperlink{A3Peres}{A3}$_\text{Peres}$ are correct,\textbf{\}} {\small THEN} \textbf{\{}the outcome of a measurement on one part of an entangled system is impacted by the apparatus setting(s) and outcome(s) of measurement(s) on other part(s) of that entangled system, even if the measurement sites are space\bz{-}like separated.\textbf{\}} 
\vspace{.5ex}\end{changemargin} 
This is a direct negation of the \al assumption of mutually independent existence of space\bz{-}like separated objects\arp\!\cite{Einstein:QMuW}, which seemed indispensable to Einstein. In a letter to Born, dated March--03--1947\cite[p.\,214\,--\,215]{BornEinstein:Briefe}, Einstein again explained that quantum theory seemed unacceptable to him, \al because the theory is not compatible with the principle, that physics shall represent a reality in time and space, without spooky actions at a distance.\ar  \hyperlink{C5Peres}{C5}$_\text{Peres}$ says that it's not quantum theory in particular, nor physics in general, but that it is the objective reality of the world in which we are living, which is not compatible with Einstein's separability principle. 

\hyperlink{C5Peres}{C5}$_\text{Peres}$ is often presented as a \al proof of quantum non\bz{-}locality\arp . That wording can be easily misunderstood, however, and needs a lot of additional explanations. Einstein's \al spooky actions at a distance\ar  became in Bell's wording \al a mechanism whereby the setting of one measuring device can influence the reading of another instrument, however remote.\arp\!\cite{Bell:Ungleichung}. Such classical\bz{-}biased characterizations of non\bz{-}separability seemed not at all appropriate to Bohr. 

In his reply to EPR, Bohr\!\cite{Bohr:EPR} pointed out that in \al the study of the phenomena of the type concerned [\dots ] we have to do with a feature of \emph{individuality} completely foreign to classical physics.\ar  Bohr emphasized the notion \emph{individuality} by italics, and he left neither in that article nor in any others of his writings any doubt, that he meant that notion literally: (Latin)\bz{\,}individual\bz{\,}=\bz{\,}not\bz{\,}divisible. Because of their individuality, quantum phenomena require a holistic description in Bohr's opinion. 

Consider as a simple classical example a thin, long rod. Let the rod be 1 lightyear long. Take one end of the rod into your hand, and point with the rod into interstellar space. The rod's center of gravity is 1/2 lightyear away from you. Now cut off \mbox{1\,m} from your end of the rod. The rod's center of gravity thereby will move \mbox{0.5\,m} from it's previous position. When will the center of gravity move? Immediately, or only 1/2 year later? Did you push it from it's previous position due to \al spooky action at a distance\arp ? Did you apply a \al mechanism\ar  whereby your cutting action did \al influence\ar  the far away center of gravity? Clearly such questions are not really sensible for a holistic property of the rod like it's center of gravity. Einstein's \al spooky action at a distance\arp , and Bell's \al mechanism\arp , with a superluminal \al signal involved\arp , completely miss the holistic character of quantum phenomena pointed out by Bohr. 

We should not adopt the classical point of view, only because it is advocated by Einstein and Bell. Nor should we adopt the holistic point of view, only because it is advocated by Bohr. Instead we should adopt the holistic point of view with regard to \emph{individual} quantum phenomena, because quantum theory is a very successful theory, which does indeed suggest the holistic approach: The state vectors \eqref{mjsdngdnsd} assign precisely defined spin projections or polarizations to the \emph{individual} \mbox{D\&G} system, but they do not assign any spin projections or polarizations to the parts D and G\,. Only due to interaction with the measurement devices, the \mbox{D\&G} system is split into the two new quantum systems D and G, and the spin projections or polarizations of D and G --- which according to \hyperlink{C3Peres}{C3}$_\text{Peres}$ did not exist before the measurement --- are created. 

An important further conclusion can be drawn from the experimental violation of the inequality \hyperlink{D4Peres}{D4}$_\text{Peres}$\,, if this additional basic assumption is made:\\\hspace*{.9\textwidth}\vspace*{-1\baselineskip} 
\vspace{.5ex}\begin{changemargin}{0em}{4.3em}\hspace{-3.7em}\hypertarget{A4Peres}{\textbf{A4}$_\text{Peres}$} Measurements have unique results.
\vspace{.5ex}\end{changemargin} 
This assumption alludes negatively to the many\bz{-}worlds interpretation proposed by Everett\!\cite{Everett:ManyWorlds} and de\,Witt\!\cite{deWitt:ManyWorlds}. In any many\bz{-}worlds interpretation of reality (not \al of quantum theory\arp ), measurements have \emph{no} unique results. Instead all possible results are realized at the same time in the various branches of reality (the \al many worlds\arp ).   

Note that the basic assumption \hyperlink{A4Peres}{A4}$_\text{Peres}$ is not needed for the conclusions \hyperlink{C1Peres}{C1}$_\text{Peres}$\bz{\,}\dots\bz{\,}\hyperlink{C5Peres}{C5}$_\text{Peres}$ listed so far. All these conclusions depend on the reality or non\bz{-}reality of results of not performed measurements, but not on the uniqueness of the results of actually performed measurements. Even \hyperlink{C4Peres}{C4}$_\text{Peres}$\bz{\,}(hidden variables do not really exist) does not depend on \hyperlink{A4Peres}{A4}$_\text{Peres}$\,: While by construction no hidden variables, which uniquely determine the results of single measurements, exist in any many\bz{-}worlds\bz{-}reality, \hyperlink{C4Peres}{C4}$_\text{Peres}$ would not be wrong but merely redundant in that framework.  

But \hyperlink{A4Peres}{A4}$_\text{Peres}$ is needed for the following consideration: Both quantum theory and classical theory are strictly deterministic. State vectors evolve unitarily according to the quantum\bz{-}theoretical equations, and classical systems evolve deterministic according to the classical equations of motion. Only in the moment of measurement something non\bz{-}deterministic is happening, as the results of measurements are not determined before the measurements, but are created only in the very moment of measurement (\hyperlink{C3Peres}{C3}$_\text{Peres}$). 

What is not pre\bz{-}determined, that can impossibly be pre\bz{-}computed. If the unique results --- which only exist if \hyperlink{A4Peres}{A4}$_\text{Peres}$ is true --- of single measurements are not pre\bz{-}determined but created only in the moment of measurement, then there definitively does not exist (and hence can never be detected) a law of nature, which rules Nature's decision for this or that particular result. If \hyperlink{A2Peres}{A2}$_\text{Peres}$ and \hyperlink{A3Peres}{A3}$_\text{Peres}$ and \hyperlink{A4Peres}{A4}$_\text{Peres}$ are correct, then the violation of Bell's inequality experimentally proves that the reason for Nature's decision is not merely unknown to us, but that there exists no reason for this or that decision. This means that nature is acting truly \emph{irrational}, when deciding for the particular measurement result: 
\vspace{.5ex}\begin{changemargin}{0em}{4.3em}\hspace{-3.7em}\hypertarget{C6Peres}{\textbf{C6}$_\text{Peres}$}\ {\small IF} \textbf{\{}\hyperlink{A2Peres}{A2}$_\text{Peres}$ and \hyperlink{A3Peres}{A3}$_\text{Peres}$ and \hyperlink{A4Peres}{A4}$_\text{Peres}$ are correct,\textbf{\}} {\small THEN} \textbf{\{}Nature is acting truly irrational (\ggie  outside the application range of any law of nature, hence outside the application range of physics) when deciding for the particular result of a single measurement.\textbf{\}} 
\vspace{.5ex}\end{changemargin} 
A `proof by contradiction' does not tell us, which of the basic assumptions is\bz{/}are wrong. Therefore the restricting condition \al if \hyperlink{A2Peres}{A2}$_\text{Peres}$ (no superdeterminism) and \hyperlink{A3Peres}{A3}$_\text{Peres}$ (no retrocausation) are correct\ar  is an indispensable part of each of the six conclusions \hyperlink{C1Peres}{C1}$_\text{Peres}$\dots\hyperlink{C6Peres}{C6}$_\text{Peres}$\,. For \hyperlink{C6Peres}{C6}$_\text{Peres}$ the additional condition \al if \hyperlink{A4Peres}{A4}$_\text{Peres}$ (measurements have unique results) is correct\ar  is essential. 

\section{Discussion}
The derivation of Bell's inequality due to Peres is much stronger than Bell's own derivation 
\vspace{-1ex}\begin{ggitemize}
\ggitem{because Peres' derivation is based onto only one \al problematic\ar  assumption, and thus the probably wrong basic assumption can easily be identified, while Bell based his derivation onto two problematic assumptions, which later both were proved wrong due to \hyperlink{C4Peres}{C4}$_\text{Peres}$ (provided that Bell assumed \al really existing\ar  hidden variables) and \hyperlink{C5Peres}{C5}$_\text{Peres}$\,.}
\ggitem{because \hyperlink{C1Bell}{C1}$_\text{Bell}$ and \hyperlink{C2Bell}{C2}$_\text{Bell}$ can be concluded from \hyperlink{C1Peres}{C1}$_\text{Peres}$\,\dots\linebreak\hyperlink{C6Peres}{C6}$_\text{Peres}$\,, while \hyperlink{C1Peres}{C1}$_\text{Peres}$\dots\hyperlink{C6Peres}{C6}$_\text{Peres}$ can impossibly be concluded from \hyperlink{C1Bell}{C1}$_\text{Bell}$ and \hyperlink{C2Bell}{C2}$_\text{Bell}$\,, \ggie because \hyperlink{C1Bell}{C1}$_\text{Bell}$ and \hyperlink{C2Bell}{C2}$_\text{Bell}$ is a proper subset of \hyperlink{C1Peres}{C1}$_\text{Peres}$\dots\hyperlink{C6Peres}{C6}$_\text{Peres}$\,.}  
\ggitem{because of the stunning simplicity of the derivation due to Peres, which may discourage attempts of refutation.}
\vspace{-1ex}\end{ggitemize}

Peres' basic assumption \hyperlink{A1Peres}{A1}$_\text{Peres}$ (\al not measured results are as real as actually measured results\arp ) and Bell's basic assumption \hyperlink{A1Bell}{A1}$_\text{Bell}$ (\al not measured results are determined by hidden variables\arp ) are similar, but not identical. Bell missed to anticipate Peres' stronger conclusions, because he was determined to proof the possibility of hidden\bz{-}variables theories.  Bell's wording \al It is the requirement of locality [\,\dots\,], that creates the essential difficulty.\ar  in the introduction of \cite{Bell:Ungleichung} indicates that he in error believed that \hyperlink{A2Bell}{A2}$_\text{Bell}$ was the only wrong one of his basic assumptions. Due to the slight, but ingenious twist from hidden variables, which hypnotized Bell, to the reality of not measured results, Peres could in the end not only proof non\bz{-}separability (\hyperlink{C5Peres}{C5}$_\text{Peres}$), but also settle the hidden\bz{-}variables stuff (\hyperlink{C4Peres}{C4}$_\text{Peres}$). 

I did not conceal my personal preferences for the assumptions of no superdeterminism, no retrocausation, and unique measurement results (\ggie  a single\bz{-}world reality). We must not forget, however, that these are merely assumptions but not proven facts. We can hardly avoid to make such assumptions in many places, and in most cases we make them tacitly, because otherwise any simple scientific statement would be burdened with a huge load of caveats. But it is advisable to keep those unproved assumptions in mind, because we might need to reassess and modify some of them in case we should get stuck in a dead end at some point of our evaluations. 

Under the premise that the assumptions of no superdeterminism, no retrocausation, and unique measurement results (\ggie  a single\bz{-}world reality) are correct, the violation of Bell's inequality experimentally proves that we do \emph{not} live in a deterministic clockwork\bz{-}universe (\hyperlink{C6Peres}{C6}$_\text{Peres}$). This is arguably the most important of the listed conclusions, as it allows for \al free will\ar  and an open future. \hyperlink{C6Peres}{C6}$_\text{Peres}$ does not prove that we are endowed with free will. \hyperlink{C6Peres}{C6}$_\text{Peres}$ merely proves that the world is not subject to brute determinism, and that free will is not excluded by the laws of nature. 

\flushleft{\interlinepenalty=100000\bibliography{../../gg}} 
\end{document}